\documentclass[english]{article}
\usepackage[utf8]{inputenc}
\usepackage[T1]{fontenc}
\usepackage{babel}
\usepackage{amsmath}
\usepackage{graphicx}
\usepackage{fancyhdr}

\usepackage[textwidth=18cm,textheight=24cm]{geometry}

\usepackage{setspace} 
\usepackage{caption,subfigure,float}

\usepackage{hyperref,url,cite}

\begin{document}

\title{FoodRepo: An Open Food Repository of Barcoded Food Products}

\author{Gianrocco Lazzari\textsuperscript{1}, Yannis Jaquet\textsuperscript{1},
	Djilani Kebaili\textsuperscript{1}, Laura Symul\textsuperscript{1}, Marcel Salathé\textsuperscript{1{*}}}

\maketitle

1. Global Health Institute, School of Life Sciences, Ecole Polytechnique Fédérale de Lausanne (EPFL), Lausanne, Switzerland.
*  corresponding author: Marcel Salathé (marcel.salathe@epfl.ch).

\begin{abstract}
	

In the past decade, digital technologies have started to profoundly influence healthcare systems. Digital self-tracking has facilitated more precise epidemiological studies, and in the field of nutritional epidemiology, mobile apps have the potential to alleviate a significant part of the journaling burden by, for example, allowing users to record their food intake via a simple scan of packaged products' barcodes.
Such studies thus rely on databases of commercialized products, their barcodes, ingredients, and nutritional values, which are not yet openly available with sufficient geographical and product coverage.
In this paper, we present FoodRepo (\href{https://www.foodrepo.org}{https://www.foodrepo.org}), an open food repository of barcoded food items, whose database is programmatically accessible through an application programming interface (API).
Furthermore, an open source license gives the appropriate rights to anyone to share and reuse FoodRepo data, including for commercial purposes.
With currently more than 21,000 items available on the Swiss market,
 our database represents a solid starting point for large-scale studies in the field of digital
 nutrition, with the aim to lead to a better understanding of the intricate connections between diets and health in general, and metabolic disorders in particular.

\end{abstract}

\section*{Background \& Summary}

Metabolic disorders, such as diabetes or obesity, have become a major public health concern, with increasingly large parts of the global population affected \cite{who_Diabetes,who_Obesity}.
Nutritional epidemiologists hope to better understand the underlying causes, the potential treatments and prevention strategies by analyzing population and individual patterns through studies that generally rely on surveying dietary habits. Traditional food-intake survey methods are based on questionnaires filled by participants at a given frequency. This frequency is crucial to determine the accuracy of the records and the scope of the research \cite{satija2015understanding}. Multiple-day diet records provide high accuracy, but require a strong motivation and time commitment by the participants. Approaches like multiple / single 24-h recalls - involving a specialized interviewer performing surveys in person or at the phone with the participants - 
	require less engagement, but pose some issues with missing data, as they rely on short-term memory. Finally, so-called Food Frequency Questionnaires (FFQ), where participants are asked to indicate the frequency of intake of certain foods over long periods of time (typically 1 year), demand minimal participants' commitment, therefore allowing for large cohort studies on long-term dietary habits. The likelihood of missing or incorrect data, however, increases as they count on participants' long-term memory. 
    
    Recent technological advances, and in particular the emergence and almost complete market penetration of smartphones, have offered interesting surveying alternatives. In particular, mobile phones have been successfully deployed in several food-related studies \cite{sharp2014feasibility}, for example using food photography \cite{chae2011volume,kong2012dietcam,lee2012comparison,dibiano2013food,zhu2011multilevel,zhu2010use}. 
	Other research has also explored the possibility of recording dietary habits by asking participants to scan the barcodes of their consumed food  \cite{siek2006we,eyles2010use}. Although further investigations are required to assess self-reporting biases, these advances in nutritional research have triggered the release of  mobile apps oriented mainly towards diabetes and weight-loss self-management \cite{pagoto2013evidence,dunford2014foodswitch,stephens2011smart,tsai2007usability,azar2013mobile}, showing the willingness and interest of users to  monitor their food intake if it provides potential health benefits. 
    
The further expansion of self-monitoring for research and medical purposes relies on comprehensive and continusouly updated food databases. A few databases of barcoded products already exist, for example Open  Food Facts \cite{open_Food_Facts} or the USDA Food Composition Databases \cite{usda_Food_Composition_Database}. While they each have their strength, not all of them are openly accessible or, and they often have a limited product coverage, and are often not regularly updated. For Swizterland, we did not find any database whose product coverage was sufficiently high, where the data was completely open, and easily accessible through an API (Application Programming Interface). The last point was particuarly important to us, as APIs are necessary for third parties to dynamically use the data in their products and services. Our approach was therefore to build an openly accessible database of barcoded food products with sufficiently high coverage, acccessible through a stable API. Rather than focusing on a wide geographic range, we focused on a small country (Switzerland) in order to obtain the necessary coverage. The focus on the Swiss market further benefits from the need to support multiple languages from the beginning, thus making the system readily expandable to other countries, which we are now planning to do.

Here, we present this system, which we call FoodRepo (\href{https://www.foodrepo.org}{https://www.foodrepo.org}), a database of barcoded food products, and we describe the data-acquisition framework, its quality control and maintenance. The growing community around FoodRepo and the validation of new products make our database robust, scalable and self-sustainable in the long run. Currently, the FoodRepo database mostly holds products sold in Switzerland, from the main grocery stores in the country. Its international expansion is under development.

\begin{figure*}
		\centering
		\includegraphics[scale=1]{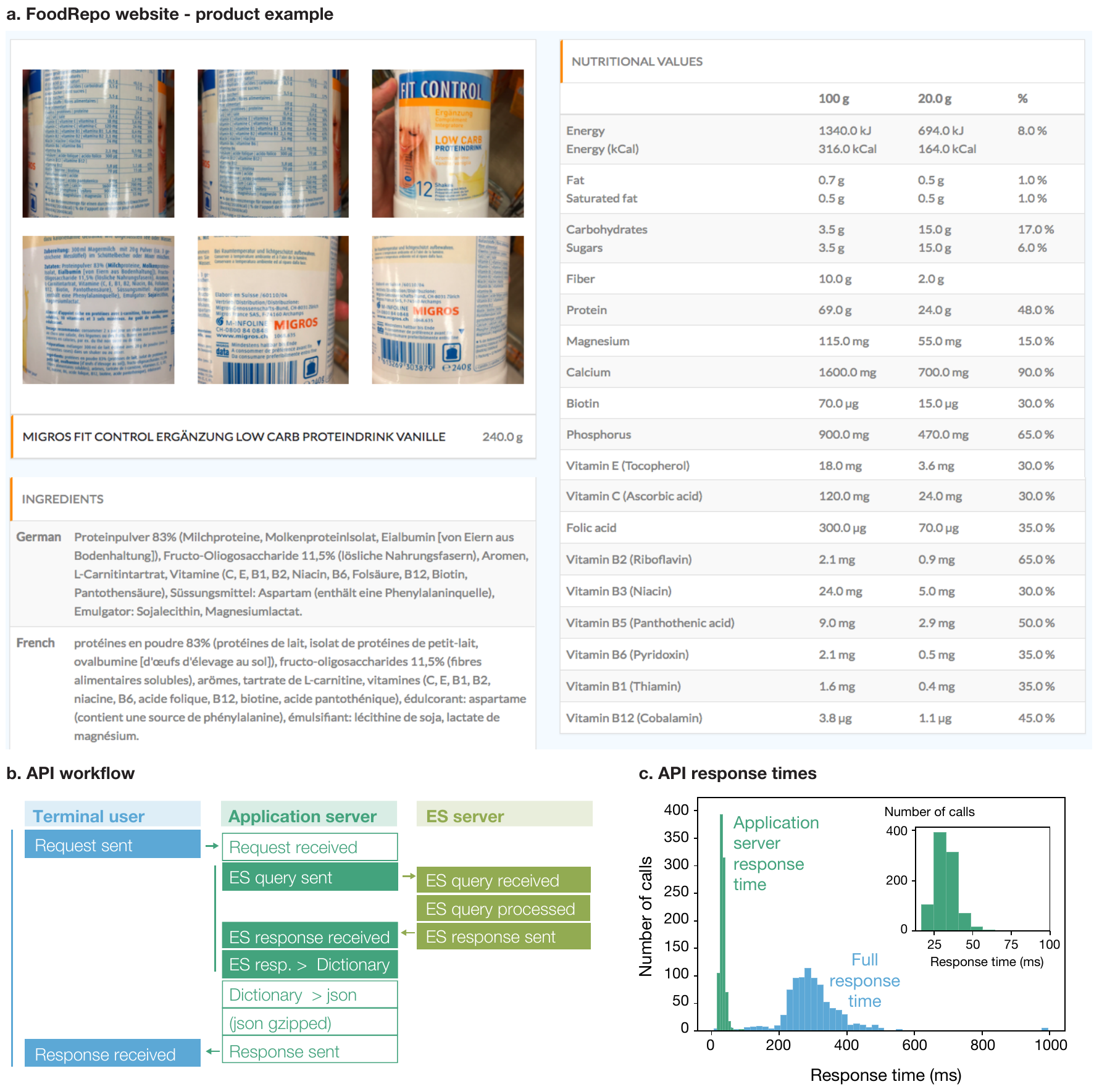}
		\caption{(a) Screenshot of the  \href{https://www.foodrepo.org}{FoodRepo} website. (b) Schematic presentation  of the pipeline behind our API. (c) Distribution of API response times, color-coded according to different sections of the back-end pipeline, as shown in panel b.}
		\label{1_fig_introduction}
	\end{figure*}

Any item in the database is accessible through the FoodRepo website (fig. \ref{1_fig_introduction}-a )
 or via our API, described in section \hyperref[usage_notes]{\texttt{Usage Notes}}. 
The CC-BY-4 license under which our database is released will allow its exploitation by different type of users, from academic researchers to commercial partners.
	For instance, a Swiss consumers association is using FoodRepo data in their NutriScan \cite{nutriscan}
to make the food package information more accessible, and to provide their users with an overall nutritional score.
   
 Beyond this specific example, the FoodRepo database opens the way for promising research opportunities in the field of digital epidemiology and personalised nutrition.
Notably, we foresee that, through dietary live-tracking, this database can support studies which combine other recent technological developments and new findings in our understanding of the human metabolism. For example, phone-connected devices for continuous monitoring of blood glucose levels have recently been made available to diabetic patients \cite{pfeiffer1989glucose,aljasem2001impact}, as well as numerous direct-to-consumer devices to estimate glucose levels have appeared on the market. A plethora of other wireless sensors are now also available to record various physiological parameters, marking a new era of `high-throughput human  phenotyping' \cite{elenko2015defining}. Studies that would simultaneously track participants' parameters, food intake, glycemic response and physical activity might provide detailed insights on the variability of individual metabolic responses. Interestingly, one of the factor which has recently been found to account for a large part of this variability is the microbiota \cite{Griffin2017,Turnbaugh2006,le2013richness,Zeevi2015,pedersen2016human}. Large-scale testing of these hypotheses through self-tracking could contribute to assess the complex metabolic response of the human body to different energy sources. 
This requires detailed records of food intake that includes nutritional information as well as eating times \cite{scheer2009adverse} and food portion sizes \cite{ello2005influence,ledikwe2005portion,young2002contribution}, challenges that FoodRepo contributes to overcome. 
   
FoodRepo will also allow researchers to compare packaged food offers across time and geographical areas. Analyses of the database evolution will give interesting indication on the dietary trends and on the overall modification of the nutritive quality of packaged food. Although the database itself does not inform on the buying frequency, the continuous introduction of specific products in the market, and thus in the database, can potentially indicate how retailers react to customer demands and changing dietary habits.

\begin{table}
	\centering
	\begin{tabular}{*{6}{|p{2cm}|p{15cm}|}}
		\hline
        \textbf{Fields}   & \textbf{Sample}    \\
        \hline
        \hline
		\textbf{Product ID} & 972  \\
        \hline
		\textbf{Barcode}    & 7611654884033 	 \\
        \hline
		\textbf{Name} 	&	Chocolat au lait 			aux noisettes \\
        \hline
		\textbf{Quantity}    & 150  \\
        \hline
		\textbf{Units}       & g  \\
        \hline
        \textbf{Portion Quantity} & 30 \\
        \hline
    	\textbf{Portion Unit}  & g \\
        \hline
        \textbf{Alcohol by Volume} & 0 \\
        \hline
		\textbf{Origin}      & Switzerland \\
        \hline
		\textbf{Ingredients}  & (FR) sucre de canne brut* (Paraguay), cacao en pâte (Pérou), pâte de noisette 4.5\% (Turquie), gousses de vanille*. Teneur en cacao du chocolat: 32\% minimum. * Ingrédients conformes aux standards du commerce équitable Fairtrade. 58.6\% du poids total. Dont sucre et produits à base de cacao avec bilan de masse. Tous les ingrédients agricoles sont issus de l'agriculture biologique 
        (DE) Rohrohrzucker* (Paraguay), karamellisierte geröstete Haselnüsse 22\% (Haselnüsse [Türkei], Rohrohrzucker [Paraguay], Wasser), Vollmilchpulver (Schweiz), Kakaobutter* (Dominikanische Republik), Kakaomasse* (Peru), Haselnusspaste 4.5\% (Türkei), Vanilleschoten*, Kakaobestandteil in der Schokoladenmasse: mindestens 32\%, * Nach Fairtrade-Standards gehandelte Zutaten. Gesamtanteil 58.6\%. Davon Kakaoerzeugnisse und Zuckerarten mit Mengenausgleich. Alle landwirtschaftlichen Zutaten stammen aus biologischem Anbau. Allergie: Enthält Haselnuss, Milch. Kann spuren von Mandeln, Soja enthalten.  \\
        \hline
		
 		\textbf{Nutrients (per 100g)}	 & 
		 Energy 		2410.0 kJ; 
 		 Energy (kCal)	577.0 kCal;	
 		 Fat 			40 g; 	
 		 Saturated fat	16 g; 	
 		 Carbohydrates 	43 g; 	
 		 Sugars			42 g;	
 		 Fiber			4 g;	
 		 Protein		10 g;	
 		 Salt			0.2 g	
 		
 		 \\
         \hline
		\textbf{Created at}   & 2016-05-31, 17:54:07  \\
        \hline
        \textbf{Updated at}   & 2017-11-16, 10:13:31  \\
        \hline
		\textbf{Pictures}     & Url to the front picture of the sample product: e.g. \href{https://d2v5oodgkvnw88.cloudfront.net/uploads_production/image/data/3941/large_myImage.jpg?v=1465828644}{ https://goo.gl/PyjjNa }    \\
		\hline
	\end{tabular}
	\caption{Sample product from the FoodRepo database with its values for the most relevant fields. While here we only provide the link to the front image of the product, an API call would provide the links to all pictures available for the requested products. A complete description of the fields provided by the API is available in the \href{https://github.com/salathegroup/openfood\_api/blob/master/v3/schema/product.md}{GitHub documentation}.}
	\label{example_table}
\end{table}



\section*{Methods}
\label{methods}


	
\begin{figure}
	\centering
	\includegraphics{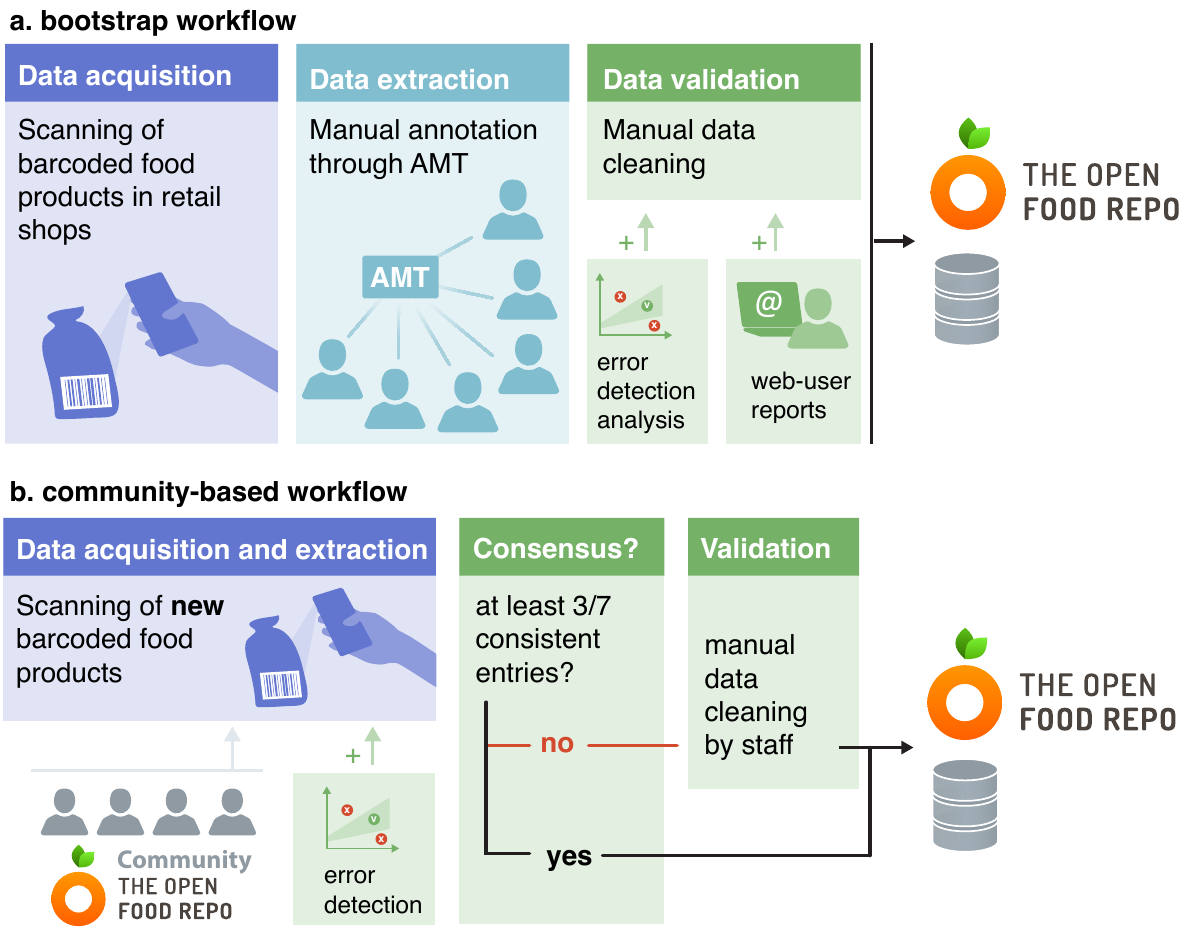}
	\caption{Schematic overview of FoodRepo data collection and validation processes.} 
	\label{of_pipeline}
	\end{figure}

    The database building and maintenance process relies on the following steps: i) collection of products pictures from local 
    retailers, ii) data extraction from the pictures, iii) validation of the extracted data,  and iv) permanent storage in the database (Fig. \ref{of_pipeline}). For the initial build of the database, we designed a specific pipeline (bootstrap workflow, Fig. \ref{of_pipeline}-a,
    which allowed us to validate the first  20,000 food products in a few months. 
    Given the dynamic nature of our data and the cost of the bootstrap workflow, we designed a second pipeline (currently under development) which relies on the growing FoodRepo community. This workflow (community-based, Fig. \ref{of_pipeline}-b) 
    allows us to keep up with the new and seasonal products introduced on the market by the retail shops, as well as to ensure the scalability and self-sustainability of FoodRepo in the long run. 

The bootstrap workflow (Fig.\ref{of_pipeline}-a)
consists of 3 main steps. 
The first step entailed a massive manual data collection from three large groceries stores in Switzerland upon approval from the shops (such as Migros, Coop, and Lidl). We hired students 
to take pictures of all barcoded food items in retail shops located in the Lausanne area. To facilitate the data collection, we specifically designed a simple phone app with which  students could scan the products' barcode and take pictures of the front and back side of the package, the product's name, ingredients list, and nutrition facts. These pictures were then automatically uploaded to the database.  At the end of this step, students had collected on average 5 pictures per item.\\
	The second step focused on the extraction of information contained in the pictures. 
	Due to the presence of multi-languages ingredients and the often wrinkled surfaces of items packages, Optical Character Recognition
	(OCR) systems could not achieve a reliable accuracy. 
	We therefore opted for a crowd-sourced solution and set up an Amazon Mechanical Turk\cite{amturk} (AMT) instance. AMT is a platform connecting \textit{requesters} to \textit{workers}, the latter being financially compensated to achieve tasks requiring human intelligence. Here, we designed a graphical user interface (GUI) allowing workers to transcribe the text they could read from products pictures. Specifically, the GUI presented text boxes where AMT workers provided the product name, nutritional values (in a table format) and ingredients, in every language present on the label (German and/or French for almost all items; Italian and/or English in addition for some products). Three different HITs (Human Intelligence Tasks) were set up: for nutrients, product name and ingredients. For the last two, we set up qualification rounds for AMT workers as their transcription involved some language skills.
	AMT workers could choose to either enter from scratch the information they saw on the pictures, or to approve / modify the suggestions given by an OCR\cite{text_recognition_API} system. 
    At the end of the second step, all annotated products were uploaded into the database, flagged as ready for validation.\\
	The third step was thus dedicated to data validation, which was based on extensive manual check by the FoodRepo team, and was additionally informed with
    manual reports by visitors to the FoodRepo website and with error-detection analyses of nutritional values. Such online reports are encouraged by the presence of a \texttt{`report an issue'} button on each product web-page, 
    which prompts a visitor to file an issue when spotting a potential error. Details about the error-detection analyses are given in the \hyperref[technical_validation]{\texttt{Technical Validation}} section. Before the final validation of the data, the FoodRepo team as well as students manually checked all products thoroughly.

	The community-based workflow (fig. \ref{of_pipeline}-b) 
	is similar to the bootstrap workflow, but instead of counting on AMT workers, it relies on the growing FoodRepo community. As new products become available in retail shops, FoodRepo users can submit them by uploading the corresponding package pictures, using the FoodRepo smartphone app. Currently, the information extraction is still performed by the FoodRepo team, but additional features are being implemented in the app, which will allow users to directly type the product details contained on the package.
    Before user-provided information is permanently stored in the FoodRepo database, consistent entries will need to be submitted by at least three different FoodRepo users. If such consensus will not be reached after seven independent submissions (i.e. there are still less than three consistent entries), the item will be manually analyzed by the FoodRepo team for definitive validation and inclusion into the database.\\
	This procedure  will ensure  minimal intervention from our team, while still guaranteeing the reliability of the data. 
	The FoodRepo team is currently fostering the development of an active community through which the continuity of FoodRepo is assured, and which will likely accelerate the birth of independent exploitations of the database, from both public and private partners.

\section*{Data Records}

%

    
	All FoodRepo data are stored in a  \href{https://www.postgresql.org}{PostgreSQL}\cite{postgreS} database, physically hosted on a server in Ireland. 
    For a quick overview of the dataset, a database dump can be downloaded from the dedicated folder in our API repository \cite{dumps_foodrepo}. However, these dumps are not generated regularly, and we strongly encourage the use of the API which delivers up-to-date information. 
    For each product, which comes with a unique numerical identifier, the database contains pictures of the item as found in the shop (usually between three to seven \texttt{.jpg} files), together with the main information presented on the package, i.e. the product name, nutritional values, ingredients list, barcode, and country of origin. The database holds as well the dates of product creation and last modification in the database (see Table \ref{example_table} ).
    The programmatic access to the database is allowed by an API, described in the section \hyperref[usage_notes]{Usage Notes}

\section*{Technical Validation}
\label{technical_validation}

\begin{figure}
		\centering
		\includegraphics[scale=1]{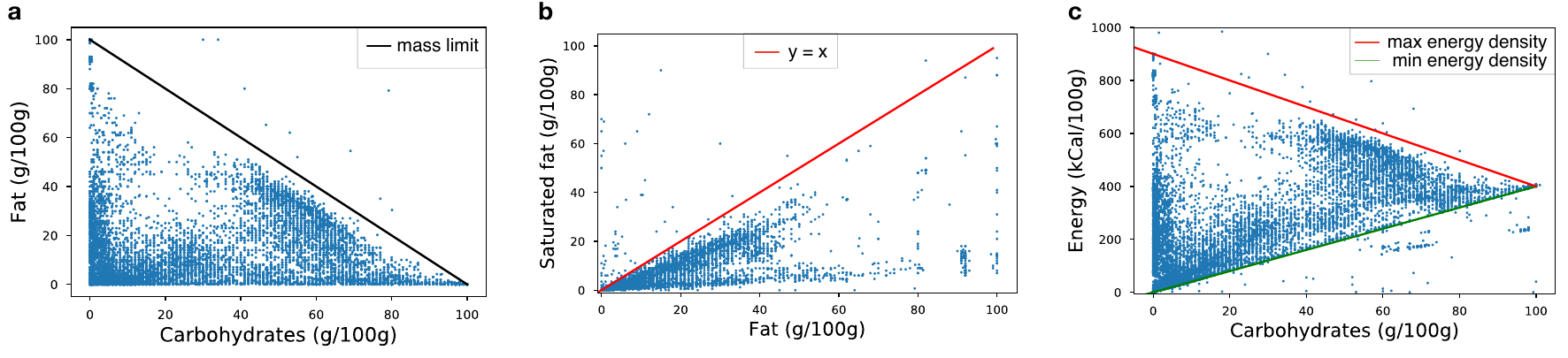}
         
		\caption{Examples of tests implemented with linear boundaries on nutritional values. Dots outside the boundaries have been inspected and corrected whenever data were different from the products packages.}
		\label{3_fig_validation}
	\end{figure}

	As described in the \hyperref[methods]{\texttt{Methods}} section, in the bootstrap stage (Fig. \ref{of_pipeline}-a) 
	the final validation was performed manually by the FoodRepo team, while in the community workflow (Fig. \ref{of_pipeline}-b),  
	the accuracy of the data is ensured by the consensus test (the FoodRepo team intervenes only if less than three matches are achieved after the uploads of the same product by seven different users). We highlight here that FoodRepo strictly reflects the information printed on products packages, even when suspicious values are present on the labels. All validation processes have thus been set-up to detect transcription errors. \\
	Within this rationale, computational analyses were  implemented for the detection of outliers, in particular regarding the nutritional values. 
	These tests reflect basic constraints, such as the mass upper-limit:
	
	\begin{equation}
	p + f + c \leq 100 
	\label{mass_boundary}
	\end{equation}
	
	where $ p,f, c  $ are the product's protein, fat and carbohydrates concentrations expressed in grams, per 100 grams of product, respectively.
	From equation \ref{mass_boundary}, one can also derive other linear inequalities for single nutrient or couples of nutrients, namely $ p + f \leq 100 $, $ p + c \leq 100 $, and $ c + f \leq 100 $.
	These simple tests allowed us to detect transcription errors in earlier versions of the database, as illustrated by the outliers in fig. \ref{3_fig_validation}-a
	which shows the distribution of products in the fat-carbohydrates space with the joint mass boundary. \\
	Similarly, other typos could be spotted by checking that the concentration of a subclass of nutrient is smaller than the one of the parent-class. This is the case for instance of sugars VS carbohydrates, or saturated-fat VS fat, shown in fig. \ref{3_fig_validation}-b. \\
	Another simple relation that helps check products' nutrition facts can be derived from the standard approximation of energy density based on nutrients composition \cite{eu_directive}:
	
	\begin{equation}
	E \sim 4p + 9f + 4c,
	\label{energy_relation}
	\end{equation}
	
	where the product's energy content $ E $ is expressed in $ kCal/ 100 \: g  $.
	Combining expressions \ref{mass_boundary} and \ref{energy_relation} provides upper and lower boundaries for the energy content (for example fig. \ref{3_fig_validation}-c).
	In this case however, not all dots that fall outside the boundaries were due to typos in transcription. 
	Indeed, the approximation in equation \ref{energy_relation} does not take into account the different contribution to energy of complex carbohydrates such as polyols, which account for less than 4 $kCal/g $. This is why products such as candies and chewing gums would fall below the energy boundaries.
			
\section*{Usage Notes}
\label{usage_notes}

%


In order to facilitate the access to the database, we built an openly accessible application programming interface (API). Any terminal user, including third party apps or services, can send API requests to retrieve specific data. The API pipeline is described in fig. \ref{1_fig_introduction}-b. User's requests are hosted on a Heroku server, where the \href{https://bonsai.io}{Bonsai Elastic Search (ES)}\cite{bonsai} application handles the queries on an \href{https://aws.amazon.com/about-aws/global-infrastructure/}{AWS server}\cite{aws} (based in Ireland). The ES response is then returned to the user after json formatting and compression (on demand). We checked that handling the request between the two servers does not critically compromise the total user-response time. We run series of single-page API calls, every 6 hours, over a week, in order to measure the full response-time and the Heroku server response-time. We observed that the latter was consistently fast across all experiments (in the range of 20-50 ms) and that the bottleneck was rather the transmission between the terminal user and the Heroku server (the average full response time was about 250 ms - see Fig. \ref{1_fig_introduction}-c).\\
For a quick introduction to the API endpoints, users are welcome to try them out on the \href{https://www.openfood.ch/api-docs/swaggers/v3}{API front page} \cite{api_Documentation}. Furthermore, on the project's GitHub repository, one can also find usage cases \cite{api_code} in Python, Ruby, Curl and JavaScript, as well as examples of complex queries \cite{api_usage_cases} in \href{https://www.elastic.co/guide/en/elasticsearch/reference/current/query-dsl.html}{ElasticSearch query DSL} \cite{queryDSL}, which include  fuzzy searches. In case of large data fetching, we suggest to use the option of compressed data\footnote{This can be done by simply setting in the request header: \texttt{Accept-Encoding: gzip}} and the possibility to include/exclude specific fields of each product (see again the API documentation \cite{api_Documentation}). In this way, one could reduce the response payload size by up to a factor of 10. \\
We remind readers that all contents (other than computer software) made available by FoodRepo on the websites, apps or services are licensed under the \href{https://creativecommons.org/licenses/by/4.0/}{Creative Commons Attribution 4.0 International License}. We however would like to highlight the fact that product images may contain copyrighted data such as brand logos.

\section*{Acknowledgements}


	This project was supported by a grant from the  \href{http://www.stiftkgj.no/?lang=en}{Jebsen foundation}\cite{jebsen_foundation}.\\
	We are grateful to Migros, Coop, and Lidl
	for access to their retail shops.

\section*{Author contributions}

	G.L. performed the descriptive and validation analysis of the dataset.
	Y.J. built the FoodRepo database, website, API and AMT HITs.
	J.D.K. maintained the API, coordinated the manual data validation and built the framework for the FoodRepo community. 
	G.L., L.S. and M.S. wrote the manuscript.
	M.S. initiated and supervised the project.

\section*{Competing financial interests}

The authors declare no competing financial interests.

\bibliography{references_foodrepo}
\bibliographystyle{ieeetr}

\end{document}